\documentclass{ws-procs9x6}

\begin{document}

\title{BARYONIC MIXING AND PRODUCTION OF HYPERNUCLEI}

\author{D.~E. LANSKOY
}

\address{Institute of Nuclear Physics, Moscow State University,
119992 Moscow, Russia\\E-mail: lanskoy@sinp.msu.ru}



\maketitle

\abstracts{ We address mixing of different baryonic states in
hypernuclei and formation of strange nuclei of various kinds.
First, production of neutron-rich $\Lambda$ hypernuclei by the
$(\pi^-,K^+)$ reaction is considered, and one-step production via
baryonic admixtures in the final as well as in the initial state
is discussed. Then, the $\Lambda\Lambda -\Xi N$ mixing in $p$
shell $\Lambda\Lambda$ hypernuclei exemplified by
$^{12}_{\Lambda\Lambda}$Be is studied. The last topic is devoted
to production of $\Theta$ pentaquark nuclei ({\it pentanuclei}).
We suggest a type of reactions with recoilless $\Theta$
production.}

\section{Production of neutron-rich $\Lambda$ hypernuclei by the
$(\pi^-,K^+)$ reaction} Interest to neutron-rich $\Lambda$
hypernuclei was recently stimulated by the KEK
experiment.\cite{Saha} In this experiment, production of
neutron-rich hypernuclei $^{10}_{\Lambda}$Li and $^{12}_\Lambda$Be
by the $(\pi^-,K^+)$ reaction was observed at the first time.

We considered\cite{TL} two mechanisms of this reaction. The first
one is the two-step production with meson charge-exchange:
$\pi^-p\to\pi^0n$, $\pi^0p\to K^+\Lambda$ or $\pi^-p\to
K^0\Lambda$, $K^0p\to K^+n$. The second mechanism is the one-step
production $\pi^-p\to K^+\Sigma^-$ proceeding via a $\Sigma^-$
doorway state. The $\Sigma^-$ admixture arises from the $\Lambda
n-\Sigma^-p$ coupling interaction. The results of our
study\cite{TL} can be summarized as follows.
\begin{itemize}
\item The cross sections of the $(\pi^-,K^+)$ reaction are
typically smaller by about three orders of magnitude than those of
the ``usual'' $(\pi^+,K^+)$ reaction.
\item The two-step mechanism is dominant.
\item Pion and kaon charge exchanges give comparable contributions.
\item The cross sections of the
$^{10}$B$(\pi^-,K^+)^{10}_\Lambda$Li reaction are higher than
those of the $^{12}$C$(\pi^-,K^+)^{12}_\Lambda$Be reaction.
\end{itemize}

The first and the last conclusions are confirmed qualitatively by
the experiment.\cite{Saha} Comparison of our
predictions\footnote{The calculated values in Table 1 are somewhat
different from those in Ref.~\refcite{TL} due to a more accurate
treatment of relevant elementary cross sections.} with the
measured data is shown in Table 1. Quantitatively the agreement is
relatively poor, which is not, however, so discouraging in view of
rather complicated nature of the process and many uncertainties in
the input information. It should be noted also that the compared
quantities in Table 1 are not of the same meaning strictly, since
the theoretical cross sections were summed over all the neutron
bound states of the final hypernuclei whereas integration over the
whole interval $0<B_\Lambda<15$ MeV has been performed in the
experiment.

\begin{table}[tbh]
\tbl{Differential cross sections of the $(\pi^-,K^+)$ reaction at
various pion incident momenta $p_\pi$ and forward angles.}
{\footnotesize
\begin{tabular}{|llcc|}
\hline reaction & $p_\pi$, GeV/c & \multicolumn{2}{c|}
{$\frac{d\sigma}{d\Omega}$, nb/sr}\\ && theory &
experiment\cite{Saha}\\ \hline
$^{10}$B$(\pi^-,K^+)^{10}_\Lambda$Li & 1.05 & 38&\\ &1.2 & 22 &
$12\pm 2$\\ $^{12}$C$(\pi^-,K^+)^{12}_\Lambda$Be & 1.05 & 5&\\
&1.2 &2.5 &7\\ \hline
\end{tabular} }
\vspace*{-13pt}
\end{table}

The main qualitative disagreement is in the energy dependence of
the cross sections. We predicted the cross sections at
$p_\pi=1.05$ GeV/c to be about twice as large as those at
$p_\pi=1.2$ GeV/c according to the sharp peak of the cross section
of the elementary $\pi N\to K\Lambda$ reaction at $p_\pi=1.05$
GeV/c. The experimental ratio is nearly inverse.\cite{Saha} The
energy dependence is still an open problem now.

Here, we would like to suggest a new mechanism of the reaction. It
is also a one-step process like production via the $\Sigma^-$
admixture but it is originated from a baryonic admixture in the
{\it initial} (nuclear) rather than in the final (hypernuclear)
state. It is known that $\Delta$ baryonic admixtures occur in
ordinary nuclei. Then, the $\pi^-\Delta^{++}\to K^+\Lambda$
process on this admixture leads just to the reaction considered.

Under usual approximations, the cross section of this process can
be estimated as
\begin{equation}
\frac{d\sigma}{d\Omega}=\frac{d\sigma_{elem}}{d\Omega}\cdot
p_{\Delta^{++}}\cdot N_{eff}.
\end{equation}
Here $\frac{d\sigma_{elem}}{d\Omega}$ is the cross section of
elementary process $\pi^-\Delta^{++}\to K^+\Lambda$. Of course, it
can not be measured experimentally. We use the theoretical
estimation\cite{Tms} based on a quark model.\footnote{This
calculation predicts a smooth energy dependence of the elementary
cross section in the relevant energy range, so the mechanism
unlikely can give an explanation of the puzzling energy dependence
of the hypernuclear production.} Then, $p_{\Delta^{++}}$ is the
probability of the $\Delta^{++}+{}^{A-1}$(Z$-$2) state, where the
nucleus is in some definite, for instance, the ground, state.
Recent experiments\cite{MB} attempted to measure $\Delta$
admixture probabilities in nuclei. From these data, we adopt
$N_{\Delta^{++}}=0.4\%$ for $^{12}$C. It should be noted that this
is the {\it total} $\Delta^{++}$ probability (not related to a
certain $^{10}$Be state), so $p_{\Delta^{++}}<N_{\Delta^{++}}$.
Thus, using $N_{\Delta^{++}}$ instead of $p_{\Delta^{++}}$, we
obtain an upper limit for the cross section.

Typical effective numbers $N_{eff}$ for specific states of final
$\Lambda$ hypernuclei are known to be (2--3)$\cdot 10^{-2}$ for
the $^{12}$C target from the theory of the $(\pi^+,K^+)$ reaction.
Assuming the same values for the reaction considered, we finally
estimate $\frac{d\sigma}{d\Omega}\sim$(10--15) nb/sr (recall that
this is an {\it upper bound}) for the
$^{12}$C$(\pi^-,K^+)^{12}_\Lambda$Be reaction at forward angles.

Comparing this value with the cross sections exhibited in Table 1,
we see that the suggested mechanism is at least non-negligible. Of
course, existence of the additional mechanism makes the problem
still more complicated. On the other hand, it might be rather
interesting if $\Delta$ admixtures in ordinary nuclei are closely
related to the production of neutron-rich $\Lambda$ hypernuclei,
and this mechanism deserves further study.

\section{Double-$\Lambda$ $p$ shell hypernuclei and
$\Lambda\Lambda-\Xi N$ mixing}

The $\Lambda\Lambda-\Xi N$ mixing in $\Lambda\Lambda$ hypernuclei
is possibly the most significant among baryonic mixings in various
nuclear systems due to the small (about 25 MeV) mass splitting
between $\Lambda\Lambda$ and $\Xi N$ pairs. Here, we study the
effect in $p$ shell hypernuclei, specifically,
$^{12}_{\Lambda\Lambda}$Be attainable by the $^{12}$C$(K^-,K^+)$
reaction. We adopt the coupled-channel model similar to that
previously applied\cite{LY} to $s$ shell hypernuclei
$^5_{\Lambda\Lambda}$H and $^5_{\Lambda\Lambda}$He.

We treat the following coupled channels. The main channel is
$^{10}$Be$_{gs}+2\Lambda$. Here, we consider $1s^2_\Lambda(L=0)$
and $1p^2_\Lambda(L=0,1,2)$ states. The second channel is
$^{11}$B$_{gs}+\Xi^-$, where $\Xi^-$ is in a $p$ state. Also the
third channel was considered, namely, $^{11}_\Xi$Z$_{gs}+N$, where
the $\Xi N$ pair possesses zero isospin, and the nucleon is in an
$s$ state. The role of the third channel is found to be typically
small if the orthogonalization of the $s$ nucleon state to the
$1s$ nucleons within $^{11}_\Xi$Z is performed. On the other hand,
the precise value of the probability of this channel depends
strongly on input parameters, so we present below the results
obtained in the two-channel approximation.

We use phenomenological Woods-Saxon potentials for
$\Lambda$-nucleus and $\Xi$-nucleus interactions. For diagonal
$\Lambda\Lambda$ and coupling $\Lambda\Lambda-\Xi N$ interactions,
potentials\cite{LY} derived from $G$ matrix calculations with
various meson-exchange Nijmegen models\cite{Nij} are employed.

\begin{figure}[tb]
\epsfxsize=56mm   
\epsfbox{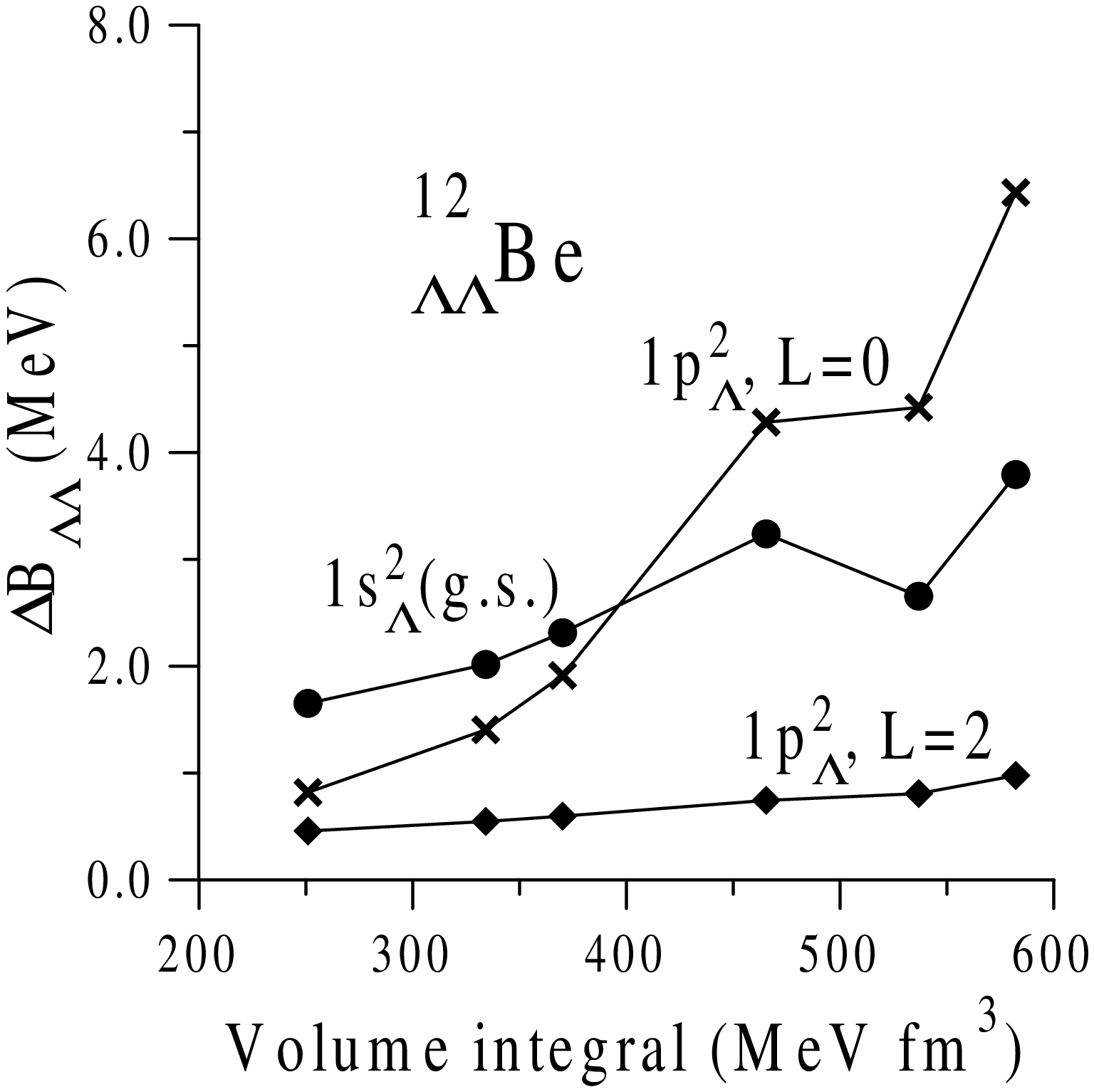}
\epsfxsize=56mm   
\epsfbox{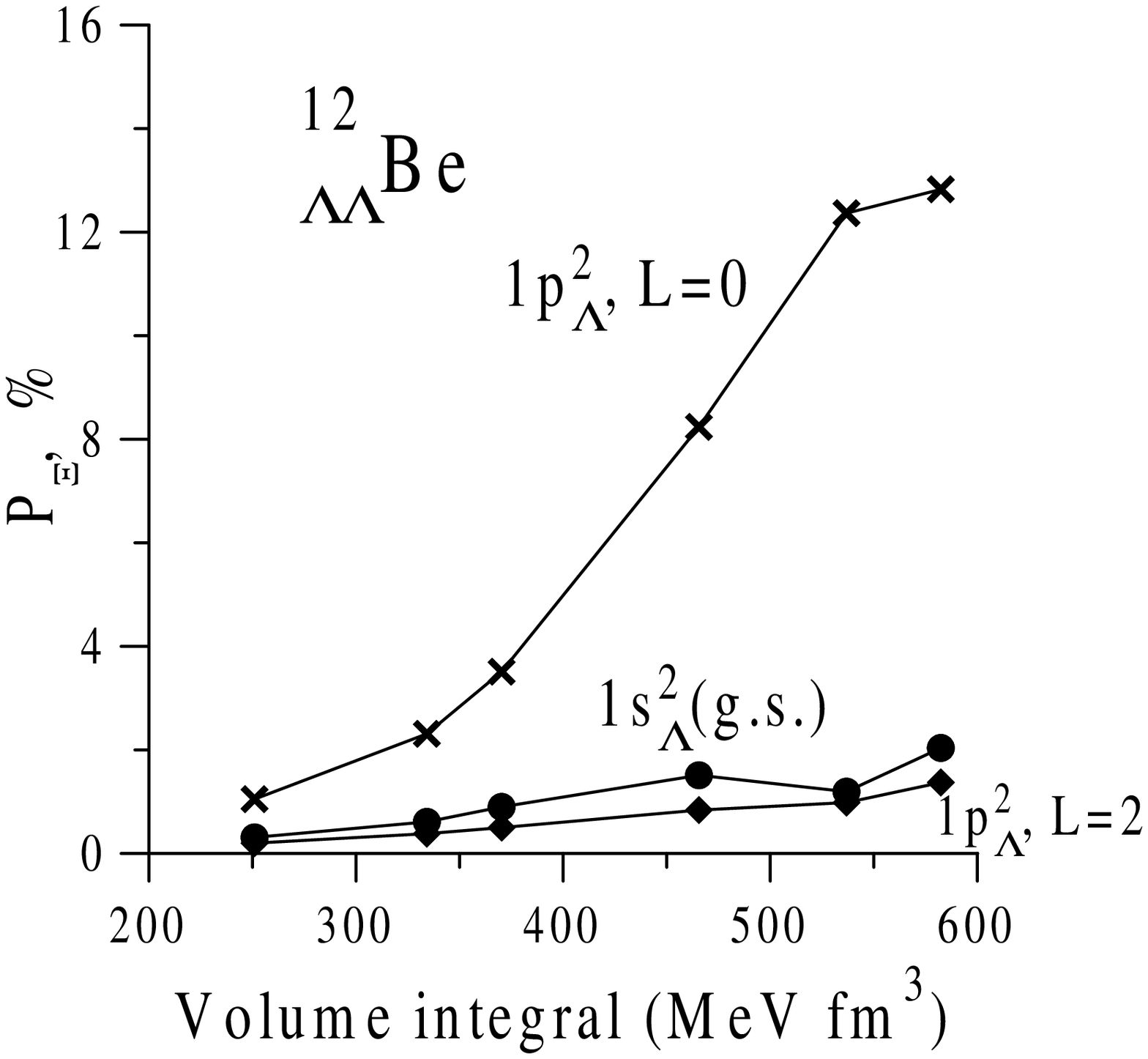}
\caption{$\Delta B_{\Lambda\Lambda}$ (top) and $\Xi$ admixture
probabilities (bottom) in $^{12}_{\Lambda\Lambda}$Be as functions
of $\int d^3rV(\Lambda\Lambda-\Xi N)$. Diagonal $\Lambda\Lambda$
and coupling $\Lambda\Lambda-\Xi N$ potentials are from Nijmegen
models\protect\cite{Nij} (from left to right) NHC-D, NSC97f,
NSC97e, NSC89, ESC03 and NHC-F.}
\end{figure}

In Fig.~1, $\Delta B_{\Lambda\Lambda}$ values (upper panel) and
$\Xi$ channel probabilities $p_\Xi$ (lower panel) are presented
for various Nijmegen models as functions of volume integral $\int
d^3rV(\Lambda\Lambda-\Xi N)$. First, it is seen that the
contribution of the coupling interaction to the energy of the
ground ($1s_\Lambda^2$) state may be considerable. For instance,
the $\Delta B_{\Lambda\Lambda}$ values for the weakest coupling
(Nijmegen hard-core D model, the leftmost point) and for the
strongest coupling (Nijmegen hard-core F model, the rightmost
point) differ by about 2 MeV. In the latter case, $\Delta
B_{\Lambda\Lambda}$ is as large as almost 4 MeV (note that all the
interactions are fitted to $\Delta
B_{\Lambda\Lambda}(^6_{\Lambda\Lambda}$He$)=1$ MeV, see
Ref.~\refcite{LY}). On the other hand, the $\Xi$ probability is
not substantially greater than 1\% anyway.

A stronger coupling effect may be expected in the $1p_\Lambda^2$
states since they lie much closer in energy to the $\Xi$
hypernuclear spectrum. It is seen in Fig.~1 that this is just the
case for the $L=S=0$ state: $\Delta
B_{\Lambda\Lambda}(1p^2_\Lambda)\equiv
B_{\Lambda\Lambda}(1p^2_\Lambda) -2B_\Lambda(1p_\Lambda)$ for
strong coupling potentials exceeds the ground state value, though
loosely bound hyperons move rather far from each other. For the
strongest coupling, $\Delta B_{\Lambda\Lambda}$ reaches a huge
value about 6 MeV. Contrary to the ground state, the $\Xi$
probabilities may be also rather great.

On the other hand, the coupling effect is relatively small for the
$L=2$ state and extremely small for the $L=1$ state (the latter is
not shown in Fig.~1). So the coupling can split levels almost
degenerated without it.

The most interesting point observed here is probably the
possibility of rather large mixing in the $1p_\Lambda^2(L=0)$
state. Due to small energy separation from the $\Xi$ hypernuclear
spectrum, excited double-$\Lambda$ hypernuclei can be essentially
states with comparable weights of the $\Lambda\Lambda$ and $\Xi N$
(and possibly also $\Sigma\Sigma$ not considered here) components.
Large $\Xi$ admixtures generally imply relatively large production
rates in the $(K^-,K^+)$ reaction via $\Xi^-$ doorway states and
the one-step $K^-p\to K^+\Xi^-$ process\cite{DGM} similar to the
mechanisms considered in Sec.~1. Note that the $\Xi^-$
probabilities obtained for $^{12}_{\Lambda\Lambda}$Be are greater
by several orders of magnitude than the $\Sigma^-$
probabilities\cite{TL} in the neutron-rich $\Lambda$ hypernuclei.
However, the production rate for the specific $1p_\Lambda^2(L=0)$
state in $^{12}_{\Lambda\Lambda}$Be is evidently small since the
$0^+\to 0^+$ transition is strongly suppressed in the $(K^-,K^+)$
reaction due to a large momentum transfer. Search for
$\Lambda\Lambda$ hypernuclear states with considerable $\Xi$
admixtures and yet producible with reasonable rates is a
challenging problem.

\section{How to inject the $\Theta$ pentaquark into a nucleus?}

The recent discovery of positive-strangeness baryon $\Theta^+$
(Ref.~\refcite{Nak}; further references are collected in
Ref.~\refcite{www}) is one of the most exciting events in
particle/nuclear physics. So far, most of experimental activities,
either performed or planned, are focused on static properties of
the pentaquark.

Several theoretical groups studied $\Theta$ interaction with
nuclei and possibility of $\Theta$ nuclei existence.\cite{ThN} By
analogy with hypernuclei, such systems may be named {\it
pentanuclei}. It is notable that, while a number of different
approaches was applied, attraction between $\Theta$ and nucleons
was deduced in all the papers,\cite{ThN} though nuclear wells vary
from very shallow to a deep one enough even to stabilize the
system with respect to strong decays.

We do not deal with any specific model of $\Theta$-nucleus
interaction. Instead, just the question formulated in the heading
of this Sec.\ is addressed.

It is well known that the key point for production of ``usual''
hypernuclei is the kinematical conditions. For instance, the main
ways to form $\Lambda$ hypernuclei are the $(K,\pi)$ reaction,
where $\Lambda$ can be produced recoillessly (with zero momentum),
as well as the $(\pi,K)$ and $(\gamma,K)$ reactions providing
$\Lambda$ momenta nonzero, but comparable to the nucleon Fermi
momentum.

Since $\Theta$ is heavier than hyperons, the momenta transferred
to $\Theta$ produced from a single nucleon ($\gamma N\to\bar
K\Theta$, $\pi N\to\bar K\Theta$, $KN\to \pi\Theta$) are rather
high (upper curves in Fig.~2). The minimal momentum transfer in
these reactions is 635 MeV/c reached in the $(K,\pi)$ reaction at
900 MeV/c.\footnote{Note that kinetic energy $T_K=300$ MeV of the
incident kaon suggested in Ref.~\refcite{Nag} as the ``optimal''
condition for the $(K^+,\pi^+)$ reaction lies well below the
threshold of the elementary reaction $K^+p\to\pi^+\Theta$, which
is about $T_K=410$ MeV ($p_K=760$ MeV/c). At $T_K=300$ MeV, the
reaction is possible only as a subthreshold many-body effect.} The
momenta transfer in the $\gamma$- and $\pi$-induced reactions (as
well as for nucleon-induced production, which is not shown in
Fig.~2) are still higher.

\begin{figure}[t]
\epsfxsize=70mm   
\epsfbox{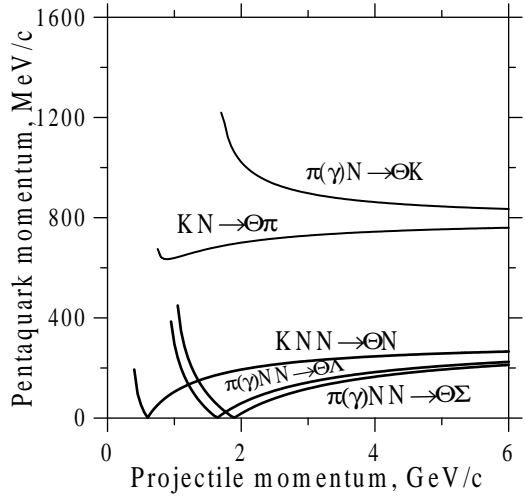}
\caption{The momentum transferred to pentaquark $\Theta$ as a
function of the projectile momentum for various reactions at zero
angle. The curves for $\gamma$- and $\pi$-induced reactions are
close to each other, therefore only the latter ones are drawn.}
\end{figure}

We suggest another type of reactions, namely, production of
$\Theta$ from a {\it nucleon pair}. It is seen from Fig.~2 that
reactions $\gamma NN\to\Theta Y$, $\pi NN\to\Theta Y$ and
$KNN\to\Theta N$, where Y is $\Lambda$ or $\Sigma$, provide really
recoilless kinematics. It is important that the momenta transfer
remain small compared to the nucleon Fermi momentum in a {\it wide
range} of the incident momentum. The relevant nuclear reactions
are as follows:
$^A$Z$(\gamma,Y){}^{A-1}_{\phantom{-1}\Theta}$Z$'$,
$^A$Z$(\pi,Y){}^{A-1}_{\phantom{-1}\Theta}$Z$'$ and
$^A$Z$(K,N){}^{A-1}_{\phantom{-1}\Theta}$Z$'$, so the reactions
are binary, like the usual reactions of hypernuclear production,
but detection of baryons (instead of mesons) is needed.

Usually cross sections of hypernuclear production are factorized
into an elementary production cross section and an effective
number involving nuclear properties. Effective numbers for the
recoilless kinematics may be expected to be greater by several
orders of magnitude than those at the momenta transfer exceeding
600 MeV/c. On the other hand, the cross sections of the elementary
(two-nucleon) processes may be small.

To get some rough estimation, we deal with pentanucleus production
on deuteron clusters and consider cross sections summed over all
final states using the closure approximation in lines of
Ref.~\refcite{DG}. We have
\begin{equation}
\frac{d\sigma_{sum}}{d\Omega}[^A\mbox{Z}(a,B)
{}^{A-1}_{\phantom{-1}\Theta}\mbox{Z}',\theta_B=0]=
\frac{d\sigma}{d\Omega}[d(a,B)\Theta,\theta_B=0]\cdot N_d\cdot
S_{dist},
\end{equation}
where $N_d$ is the deuteron cluster number taken from a cluster
model\cite{KR}, $S_{dist}$ is the distortion factor derived using
the eikonal approximation.

Any experimental information on the reactions $ad\to B\Theta$,
where $B$ is a baryon, is still absent. The reaction $\gamma
d\to\Theta Y$ was considered theoretically.\cite{Guz} So we
performed estimations for the
$^{12}$C$(\gamma,\Lambda){}^{11}_\Theta$C reaction.

Using the cross sections from Ref.~\refcite{Guz}, we estimate the
summed cross section as follows:
$\frac{d\sigma_{sum}}{d\Omega}[{}^{12}\mbox{C}(\gamma,
\Lambda){}^{11}_\Theta\mbox{C},\theta_\Lambda=0]= 29$, 10 and 2
nb/sr at $E_\gamma=1.2$, 1.6 and 2.0 GeV, respectively. Such cross
sections are evidently measurable at existing facilities. It
should be noted, however, that the elementary cross sections in
the model\cite{Guz} are proportional to the unknown $\Theta$
width, and the calculations\cite{Guz} have been performed at an
arbitrary value of 5 MeV. If the $\Theta$ width is about 1 MeV as
argued recently,\cite{Cahn} the cross sections must be reduced by
the factor of 5.

The summed cross section is not directly related to production of
low-lying pentanuclear states, but gives rather an upper bound.
From the theory of the $^A$Z$(K^-,\pi^-){}^A_\Lambda$Z reaction
(also recoilless) it has been known\cite{Boy} that about a half of
the summed cross section is contributed by low-lying
substitutional states for the $^{12}$C target. So one may expect
that the fraction of the low-lying pentanuclear states in the
summed cross section is also substantial.

Of course, it is rather problematic now to predict the elementary
cross sections reliably, so the experimental data on the $ad\to
B\Theta$ reactions are strongly needed. The g10 experiment at
TJNAF, in which the $\gamma d\to\Lambda\Theta$ reaction is
studied,\cite{CEBAF} is encouraging from this point of view. The
reaction $\gamma+{}^3$He with $\Lambda$ emission\cite{CEBAF} is
also rather interesting since {\it pentadeuteron} $^2_\Theta$He
can be produced, if it exists. On the other hand, the
meson-induced reactions deserve study as well.

To conclude this section, the pentaquark production from a nucleon
pair is probably the most promising way for formation of $\Theta$
nuclei (pentanuclei).

\section*{Acknowledgements}
I am grateful to my collaborators T.~Yu.~Tretyakova and
Y.~Yamamoto involved in some parts of this study. The work was
supported partially by Russian National Program for leading
scientific schools, Grant 1619.2003.2.


\begin{thebibliography}{99}
\bibitem{Saha} P. K. Saha, HYP2003, to be published in {\it Nucl. Phys. A};
T. Fukuda, DA$\Phi$NE2004 and these Proceedings.

\bibitem{TL} T. Yu. Tretyakova and D. E. Lanskoy, {\it Nucl. Phys.}
{\bf A691}, 51c (2001); {\it Phys. At. Nucl.} {\bf 66}, 1651
(2003).

\bibitem{Tms} K. Tsushima, S. W. Huang and A. Faessler, {\it J. Phys.}
{\bf G21}, 33 (1995); {\it Aust. J. Phys.} {\bf 50}, 35 (1997).

\bibitem{MB} C. L. Morris et al.,
{\it Phys. Lett.} {\bf B419}, 25 (1998); V. M. Bystritsky et al.,
{\it Nucl. Phys.} {\bf A705}, 55 (2002).

\bibitem{LY} D. E. Lanskoy and Y. Yamamoto,
{\it Phys. Rev.} {\bf C69}, 014303 (2004).

\bibitem{Nij} M. M. Nagels, T. A. Rijken and J. J. de Swart,
{\it Phys. Rev.} {\bf D15}, 2547 (1977); P. M. M. Maessen, Th. A.
Rijken and J. J. de Swart, {\it Phys. Rev.} {\bf C40}, 2226
(1989); Th. A. Rijken, V. G. J. Stoks and Y. Yamamoto, {\it Phys.
Rev.} {\bf C59}, 21 (1999); V. G. J. Stoks and Th. A. Rijken, {\it
Phys. Rev.} {\bf C59}, 3009 (1999).

\bibitem{DGM} C. B. Dover, A. Gal and D. J. Millener, {\it Nucl. Phys.}
{\bf A572}, 85 (1994).

\bibitem{Nak} T. Nakano et al.,
{\it Phys. Rev. Lett.} {\bf 91}, 012002 (2003).

\bibitem{www} T. Hyodo,
http://www.rcnp.osaka-u.ac.jp/\verb"~"hyodo/research/Thetapub.html

\bibitem{ThN} G. A. Miller, {\it Phys. Rev.} {\bf C70}, 022202
(2004); H.-C. Kim, C.-H. Lee and H.-J. Lee, hep-ph/0402141; D.
Cabrera et al., nucl-th/0407007; X. H. Zhong et al.,
nucl-th/0408046; F. S. Navarra, M. Nielsen and K. Tsushima,
nucl-th/0408072; V. B. Kopeliovich and A. M. Shunderuk,
nucl-th/0409010; M. J. Vicente Vacas et al., nucl-th/0410056; H.
Shen and H. Toki, nucl-th/0410072.

\bibitem{Nag} H. Nagahiro et al., nucl-th/0408002.

\bibitem{DG}  R. H. Dalitz and A. Gal, {\it Phys. Lett.}
{\bf B64}, 154 (1976).

\bibitem{KR} V. G. Kadmensky and Yu. L. Ratis, {\it Yad. Fiz.}
{\bf 33}, 911 (1981).

\bibitem{Guz} V. Guzey, {\it Phys. Rev.} {\bf C69}, 065203 (2004).

\bibitem{Cahn} R. N. Cahn and G. H. Trilling, {\it Phys. Rev.} {\bf D69},
011501R (2004).

\bibitem{Boy} A. Bouyssy, {\it Nucl. Phys.} {\bf A290}, 324 (1977).

\bibitem{CEBAF} V. D. Burkert, R. de Vita and S. Niccolai,
nucl-ex/0408019.
\end{thebibliography}
\end{document}